# Stability of Superconducting $Nd_{0.8}Sr_{0.2}NiO_2$ Thin Films

Xiang Ding[1#], Shengchun Shen[2#], Huaqian Leng[1], Minghui Xu[1], Yan Zhao[1], Junrui Zhao[1], Xuelei Sui[3], Xiaoqiang Wu[4], Haiyan Xiao[1], Xiaotao Zu[1], Bing Huang[3], Huiqian Luo[5,6], Pu Yu[2], and Liang Qiao[1*]

[1]*School of Physics, University of Electronic Science and Technology of China, Chengdu 610054, China*

[2]*Department of Physics, Tsinghua University, Beijing 100084, China.*

[3]*Beijing Computational Science Research Center, Beijing, 100193, China*

[4]*School of Mechanical Engineering, Chengdu University, Chengdu, 610106, China*

[5]*Beijing National Laboratory for Condensed Matter Physics, Institute of Physics, Chinese Academy of Sciences, Beijing 100190*

[6]*Songshan Lake Materials Laboratory, Dongguan, Guangdong 523808, China*

# Abstract

The discovery of superconducting states in the nickelate thin film with infinite-layer structure has paved a new way for studying unconventional superconductivity. So far, research in this field is still very limited due to difficulties in sample preparation. Here we report on the successful preparation of superconducting $Nd_{0.8}Sr_{0.2}NiO_2$ thin film ($T_c$ = 8.0 ~ 11.1 K) and study the stability of such films in ambient environment, water and under electrochemical conditions. Our work demonstrates that the superconducting state of $Nd_{0.8}Sr_{0.2}NiO_2$ is remarkably stable, which can last for at least 47 days continuous exposure to air at 20 °C and 35% relative humidity. Further we show the superconductivity disappears after being immersed in de-ionized water at room temperature for 5 hours. Surprisingly, it can also survive under ionic liquid gating conditions with applied voltage up to 4 V, which is even more stable than conventional perovskite complex oxides.

Supported by NSFC of China (Grant No. 11774044, 52072059, 11822411).

#These authors contributed equally to this work.

*Corresponding author. Email: liang.qiao@uestc.edu.cn

# 1 Introduction

The discovery of superconductivity in epitaxial nickelate film by Li et al.[1] has triggered a wide range of interest recently due to the delicate position of nickel in the periodic table, i.e. between copper and iron. Therefore, superconducting nickelates are expected to establish a connection between the copper and iron-based high-temperature superconductors (HTS) and provide clue for understanding the pairing mechanism in HTS [1-3]. Although showing many similarities with cuprates, for example, infinite layer structure, $3d^9$ configuration and even similar phase diagram [4, 5], many fundamental issues remain to be unraveled in nickelates. In comparison to cuprate and pnictide HTS[6-10], it shows much lower onset superconducting transition temperature ($T_c$) [11], weaker insulating characters outside the superconducting dome [4], Mott-Hubbard-type band structure [12, 13], unclear magnetic structure/interactions and absence of superconductivity in bulk materials [14-16]. Unlike cuprates and other oxide superconductors that can be easily synthesized through conventional ceramic or film synthesis techniques, the superconducting nickelate film seems to be rather vulnerable due to its narrow growth window and unclear topotactical reduction mechanism.

In fact, the removal of apical oxygen in perovskite nickelates is crucial to achieve two-dimensional $NiO_2$ planes based superconducting infinite-layer nickelates. The pioneering work of utilizing $CaH_2$ to reduce the perovskite structure to obtain an infinite-layer structure was reported as early as 1983 [17]. Yet, superconductivity in nickelate is still in its infancy stage. Due to our limited understanding of the secondary phase formation and underlying mechanism of the reduction process, effective synthesis control of infinite-layer nickelate superconductor appears to be extremely difficult. Since the first discovery of nickel-based superconductor in 2019, nearly two hundred related articles have been published while most of these are theoretical studies and experimental studies only account for a small part [1, 4, 5, 11-14, 18-28]. Particularly, there are only a few research groups worldwide that can reproduce the superconducting nickelate epitaxial films [1, 5, 19, 23, 26, 29].

Recently, significant progresses on experimental studies have been demonstrated. Both Li et al. [4] and Zeng et al. [5] have independently reported Sr substitution dependent superconducting phase diagram, and identified superconducting dome with optimized Sr concentration between 12% ~ 25 %, which is in analogy to the well-known superconducting dome for cuprates. Goodge et al. [13] employed state-of-art electron microscopy along with electron energy-loss spectroscopy to probe the Mott-Hubbard character of $Nd_{1-x}Sr_xNiO_2$ and observed emergent hybridization reminiscent of the Zhang–Rice singlet, providing direct evidence for the multiband electronic structure of the superconducting infinite-layer nickelates. Charge density waves have been discovered in both undoped and less Sr doped $NdNiO_2$ and $LaNiO_2$ films [30-32]. There are some controversies on the magnetic ground states of superconducting nickelates. Zeng et al. had used superconducting quantum interference device (SQUID) to measure the magnetic properties of superconducting and determined a perfect diamagnetism [33]. However, Lu et al. studied both pristine and doped $NdNiO_2$ by resonant inelastic X-ray scattering (RIXS) and found interesting low-energy magnon excitations with antiferromagnetic correlations [27]. Very recently, some preliminary nuclear magnetic resonance (NMR) had been done on the infinite-layer nickelates, for example, Zhao et al. [34] found a paramagnetic ground state in LaNiO$_2$ ($^{139}$La NMR) and Cui et al. [18] observed a short-range antiferromagnetic ordering by proton-enriched Nd$_{0.85}$Sr$_{0.15}$NiO$_2$ ($^1$H NMR). These NMR samples are all powder samples without superconductivity and the results are yet to be fully understood.

Despite important breakthroughs, many basic properties are still unclear in this new superconducting system, such as efficient preparation, environment stability, carrier concentration controllability of these infinite-layer nickelates, etc. In particular, the stability of superconductivity in infinite nickelates is crucial for future applications. To this end, we report the successful synthesis of Nd$_{0.8}$Sr$_{0.2}$NiO$_2$ thin films and study its environmental stability. Epitaxial Nd$_{0.8}$Sr$_{0.2}$NiO$_2$ are prepared by using pulsed laser deposition (PLD) and successive topochemical reduction method, showing highest superconducting transition temperature ($T_c$) ~ 11.1 K. The superconductivity is found to be rather stable even exposed in the air at 20 °C and 35% relative humidity (RH) for at least 47 days. However, sample soaked in de-ionized water at room temperature lost its superconductivity in 5 hours. Additionally, the superconducting nickelate film shows high energy barrier for evolution of hydrogen and oxygen ions evidenced by in-situ ionic liquid gating (ILG) [35-38], suggesting the stability against strong electrochemical conditions.

# 2 Experimental methods

The $Nd_{0.8}Sr_{0.2}NiO_2$ films with infinite layer structure are prepared by topochemical reduction of perovskite $Nd_{0.8}Sr_{0.2}NiO_3$ without capping layer. $Nd_{0.8}Sr_{0.2}NiO_3$ films (thickness about 14.5-16 nm) are deposited on $TiO_2$-terminated STO (001) substrates by 248-nm KrF laser [39, 40]. During the deposition, the substrate temperature is controlled at 620 °C with the Oxygen pressure of 200 mtorr. A laser fluence of $1J/cm^2$ was used to ablate the target and the size of laser spot is about 3 $mm^2$. After deposition, the samples were cooled down in the same oxygen pressure at the rate of 10 °C/min. In order to acquire the infinite-layer nickelate phase, the as-grown samples were cut into four pieces and sealed in the quartz tube together with 0.1g $CaH_2$. The pressure of the tube is about 0.3 mtorr [41]. After that, the tube was heat up to 300 °C in tube furnace and hold for 2 hours before naturally cooled down, with the ramp rate of 10 °C/min. The crystal structure of films is characterized using a Bruker D8 Discover diffractometer in both diffraction and reflectance mode. A hybrid monochromator, consisting of four-bounce double-crystal Ge (220) and a Cu x-ray mirror, was placed in the incident beam path to generate monochromated Cu $K\alpha_1$ X rays ($\lambda =1.54$ Å) with a beam divergence of 12 arc sec and angular precision of 0.0001°. Temperature dependent resistivity and Hall effect were measured using both four-probe method and van der Pauw geometry in a cryogen-free magnet system (CFMS, Cryogenic Ltd.) and physical properties measurement system (PPMS, Quantum Design Inc.) with magnetic fields up to 9 T and temperature down to 1.6 K. Electrical contact of Al wires were bond on deposited Au pads on the film surface by ultrasonic wire bonder. For in-situ electrical transport in ILG, the sample was placed in a quartz bowl covered entirely with ionic liquid and a slice of Pt was used as the gate electrode. The gate voltage ($V_G$) was changed at room temperature with a dwell time of 10 minutes for each cycle.

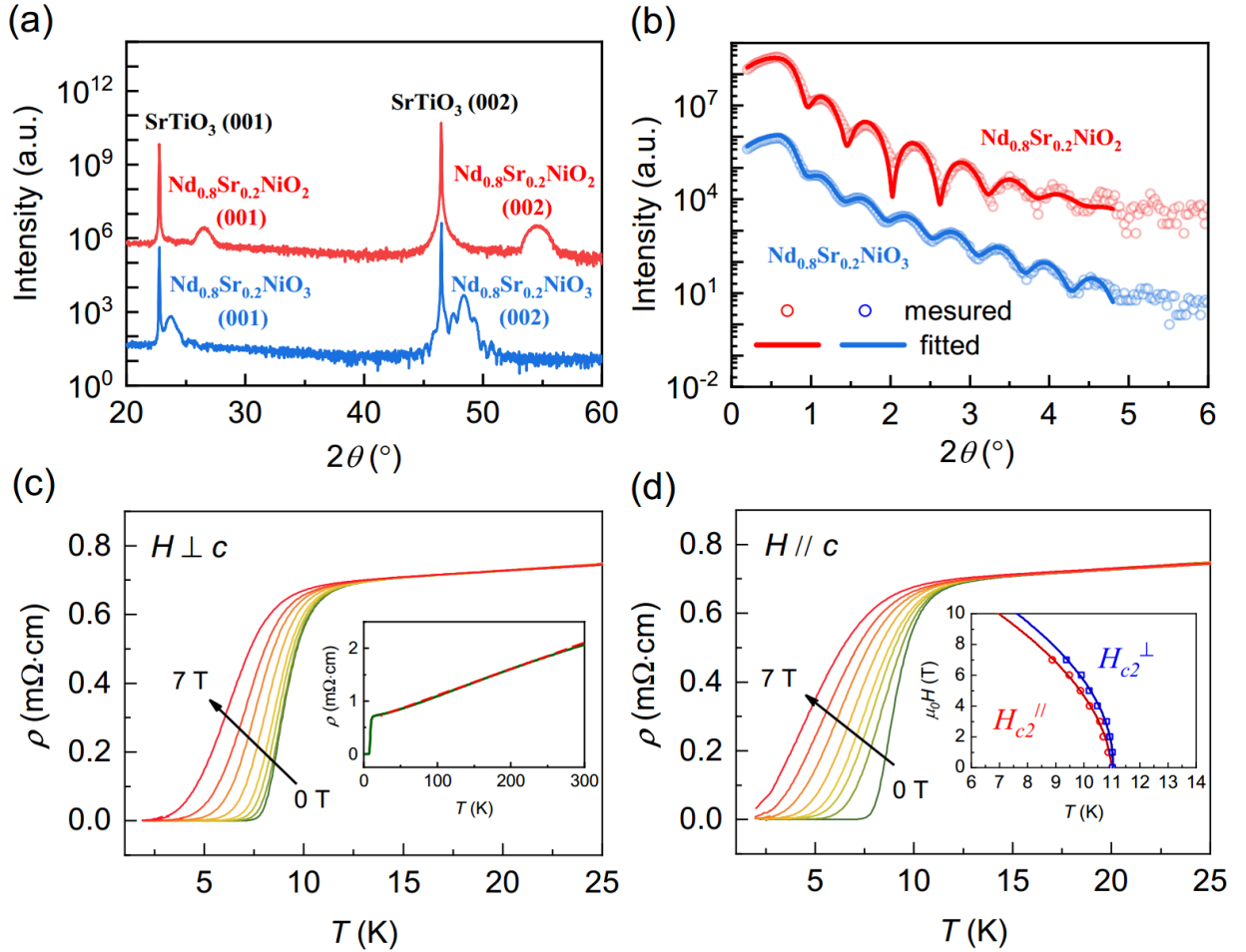

**Figure 1** (a) XRD $2\theta$-$\omega$ scans of typical $Nd_{0.8}Sr_{0.2}NiO_3$ and $Nd_{0.8}Sr_{0.2}NiO_2$ films with a thickness about 15nm. (b) Measured along with fitted XRR of as-grown $Nd_{0.8}Sr_{0.2}NiO_3$ and reduced $Nd_{0.8}Sr_{0.2}NiO_2$. $\rho(T)$ of $Nd_{0.8}Sr_{0.2}NiO_2$ film with magnetic field perpendicular (c) and parallel (d) to the $c$ axis. The insert of (c) shows the resistivity variation at all temperature and the dashed red line is the linear fitting of normal state resistivity curve. The insert of (d) shows $H_{c2}$–$T$ phase diagrams for magnetic fields along and perpendicular to the $c$ axis. The upper critical field is obtained using 90% resistivity criterion.

## 3 Result and discussion

Figure 1(a) shows x-ray diffraction (XRD) $2\theta$-$\omega$ scans of as-grown

$Nd_{0.8}Sr_{0.2}NiO_3$ and reduced $Nd_{0.8}Sr_{0.2}NiO_2$. After reduction, the film peaks shift from low to high $2\theta$ values, revealing the successful removal of apical oxygen atoms in perovskite $Nd_{0.8}Sr_{0.2}NiO_3$ and realization of infinite-layer $Nd_{0.8}Sr_{0.2}NiO_2$. The determined *c*-axis lattice constant shrinks from 3.76 Å to 3.36 Å (nearly by 11% reduction), consistent with prior work [42]. Experimental X-ray reflectivity (XRR) data of films (before and after $CaH_2$ reduction) as well as corresponding fitting curves (based on the *LEPTOS 7.10* software [43]) are displayed in Figure 1(b). Kiessig fringes are clear in the XRR curves for both perovskite and infinite-layer structures, which provide another way to determine the thickness of epitaxial thin film. The fitted thickness before and after the reduction are 14.7 and 13.8 nm, respectively. Therefore, the film thickness shrinks only by 6%, which is less than the nominal 11% shrinkage value of the lattice constant according to diffraction peak position. The discrepancy reflects there must be defect formation, which is associated with the Ruddlesden-Popper (RP)-type faults, as reported in various superconducting infinite-layer nickelates systems by transmission electron microscopy [11, 24, 44].

Figure 1(c) and (d) shows the in-plane resistivity $\rho_{ab}(T)$ of $Nd_{0.8}Sr_{0.2}NiO_2$ film at different magnetic field with both in-plane (IP, $H\perp c$) and out-of-plane (OOP, $H // c$) directions. Under zero-field condition, it is seen that the pristine film shows obvious onset (offset) $T_c$ at 11.1 K (8.2 K), where onset (offset) is defined to be the temperatures at which the resistivity is 90% (10%) of the resistivity value at 20 K. In the high temperature region (T > 50 K), $\rho_{ab}(T)$ shows a nearly linear behaviors in the normal states, as shown by the dashed line in the insert of Figure 1(c). This is also consistent with previous understanding of metallic nickelate, and its unconventional nature with possible non-Fermi liquid (NFL) state. In principle, the in-plane resistivity $\rho_{ab}(T)$ can be fitted by a power law: $\rho_{ab}(T) = \alpha T^{n}+\rho_0$, where $\rho_0$ is the residual resistivity, and $\alpha$ is a measure of the strength of electron-phonon scattering. For a Fermi liquid system, $n = 2$, thus $n < 2$ usually means a non-Fermi liquid behavior. In this case, $n = 1$ indicates possible quantum criticality in this system. Such linear $\rho_{ab}(T)$ in the normal state is a hallmark of abnormal behaviors in cuprates. In our superconducting films, we find the $\alpha$, $\rho_0$ and residual resistance ratio ($RRR=\rho(300$ K$)/\rho_0$) are sample sensitive. For example, $\alpha$ varies from 1.2 to 5.8 μΩ·cm/K, $\rho_0$ locates between 0.18 and 0.8 mΩ·cm, whereas, the $RRR$ value varies from 3.2 to 4.4. These results are consistent with the reported values in literatures [1, 19, 20]. The changes of these parameters reflect the variation of film crystal quality. This may originate from the fact that the competing phases of RP faults and perovskite during epitaxial growth show rather comparative formation energies, thus, they always coexist. The critical role of the defects on carrier transport for superconducting nickelates deserves further systematically investigation. Meanwhile, transport measurement under external magnetic field shows similar broadening tendency of suppressed superconductivity in both IP and OOP magnetic field. The $H_{c2}$–$T$ phase diagrams in Figure 1(d) indicates largely isotropic behavior in different field direction, revealing the superconductivity in nickelates has 3D characteristics. These observations are generally consistent with the reported transport measurements [20, 24].

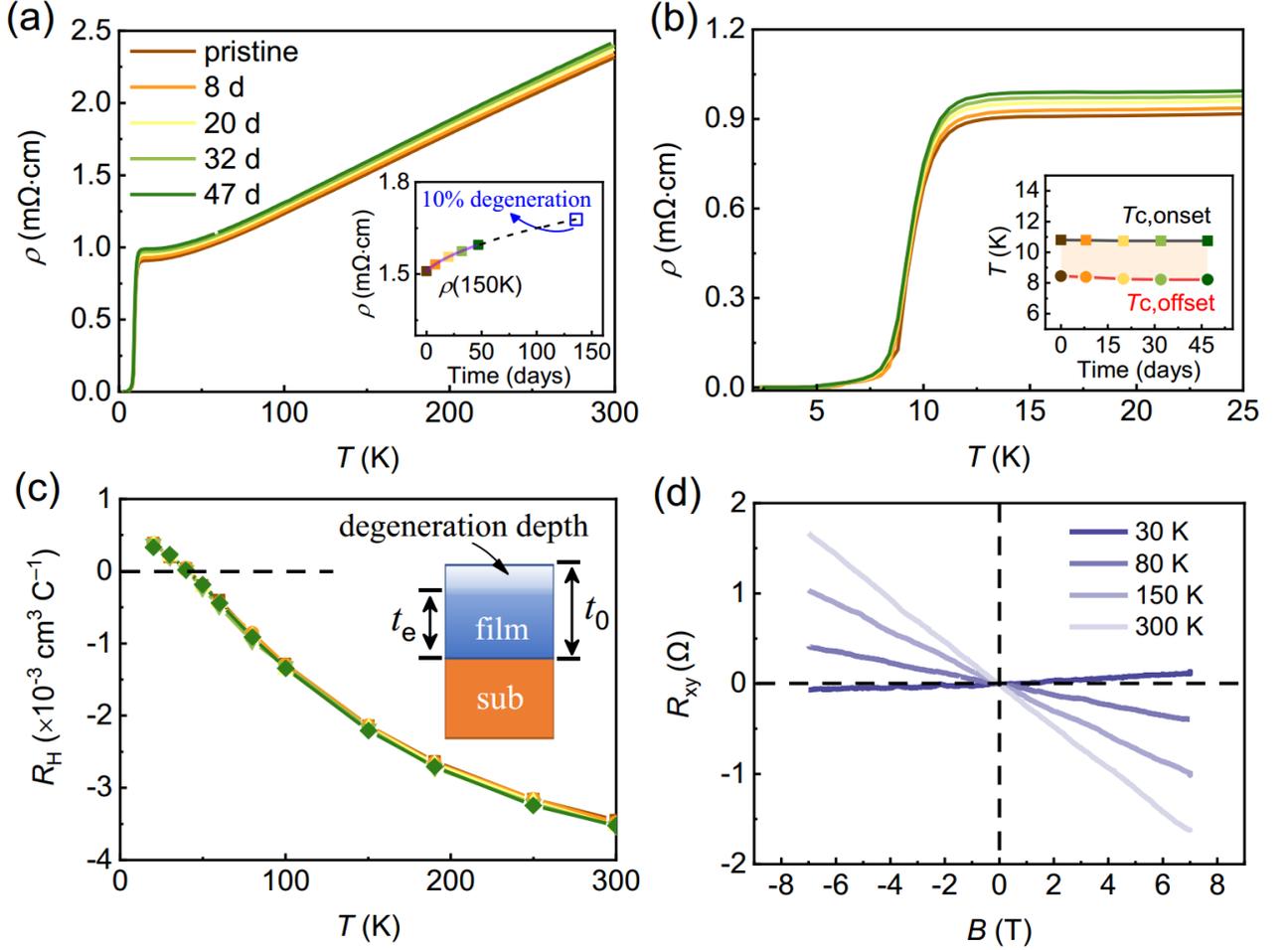

**Figure 2** (a), (b) $\rho(T)$ of a superconducting $Nd_{0.8}Sr_{0.2}NiO_2$ film continuously exposed to air for weeks. Insert of (a) shows the $\rho(150\ K)$ versus exposure time (solid squares) and the fitting curve. The blue hollow square indicates the film with predicted 10% degeneration. Insert of (b) shows superconducting transition temperatures. (c) Normal-state Hall coefficient $R_H(T)$ corresponding to (a). Insert illustrates the surface degeneration model. (d) Linear behavior of Hall resistances $R_{xy}$ in superconducting film.

After confirming the basic physical properties of the superconducting film, we began to study the stability of the film. As far as we know, many of the copper-oxide superconductors degrade rapidly in the presence of air, which contains oxygen, carbon dioxide, moisture, etc. [45] Among published experiments, most superconducting samples are stored in vacuum chamber or even in glove box under nitrogen atmosphere to avoid the contact with oxygen and moisture in the air [19, 20]. However, the stability of infinite nickelate in air condition is still unclear. Therefore, it is important to evaluate the stability of infinite-layer nickelates in this condition. To answer the question, a superconducting sample without capping layer was prepared and exposed in the air at ambient condition (20 °C, 35% RH) to measure its $\rho(T)$ curve and temperature-dependent normal-state Hall coefficient ($R_H$) every several days up to a total of 47 days. As displayed in Figure 2(a), except for a very slight increase of normal-state resistivity over time, the superconductivity hardly changes. Interestingly, we found that the increase in resistivity is not linear but has a power relationship with exposure time ($t$), i.e. $\rho(t) \propto t^{0.5}$. As a proxy for this law, resistivity versus exposure time at 150 K ($\rho$(150 K)) is shown in the insert of Figure 2(a). Details in superconducting transition regime are illustrated in Figure 2(b). Although the resistance increases monotonically with time, the change in $T_c$ is very slightly (see Figure 2(b) insert).

Figure 2(c) shows the temperature-dependent $R_H$ corresponding to Figure 2(a) for films exposure to air up to 47 days. The linear behavior of $R_{xy}$ up to 7 T at different temperatures enable us to efficiently extract $R_H$ (Figure 2(d)). It is obvious that $R_H$ has a strong temperature dependence. At room temperature, the $R_H$ has a negative sign and large magnitude, which is attributed to negatively-charged electrons with low concentration. As the temperature decreases, the carrier concentration gradually increases, and finally transitions to hole-type carriers (< 40 K). The above analysis of Hall data is based on simple carrier density ($n = 1/eR_H$, where $e$ is the electron charge) estimation. Generally, $Nd_{0.8}Sr_{0.2}NiO_2$ is expected to be a hole-type superconductor due to divalent Sr-doping on the rare-earth position, thus the Hall coefficient should be positive. However, due to the complex band structure and the self-doping effect from Nd element (detailed explanation can be found in the following discussion), *n*-type behavior can also be observed. As a result, through Hall measurement, both positive and negative Hall coefficient are determined. Although the exact physical origin of both type of carriers is not clear, our data is consistent with literature reports [4, 5]. It is evidenced that the Hall coefficient of the sample in an air environment is very stable, except for the Hall coefficient at room temperature deviates a little to pristine curve. As the temperature decreases, the deviation gradually decreases. Therefore, our data suggest that superconducting nickelates are stable in conventional ambient conditions (dry air) and glove box is not necessary for sample storage.

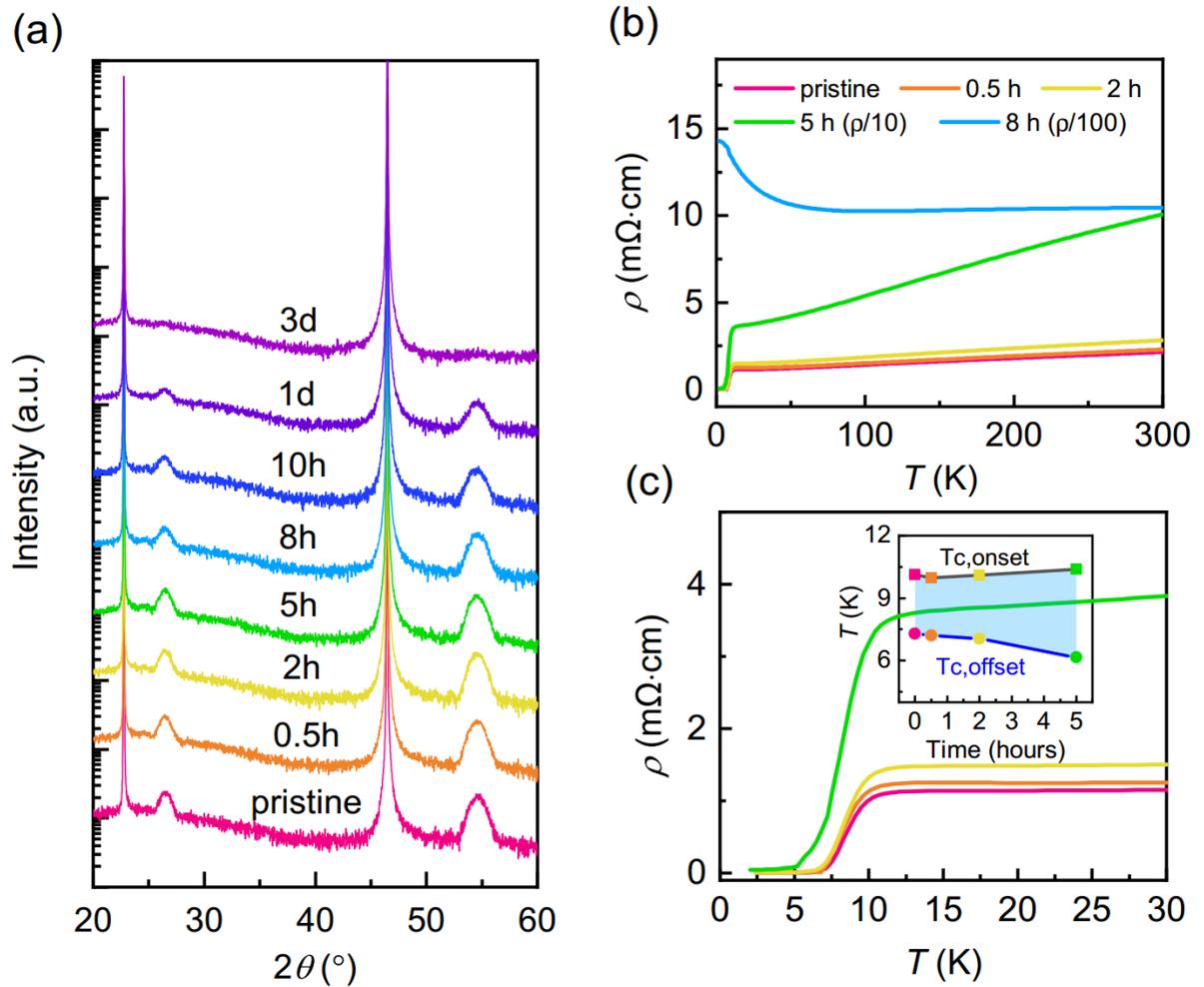

**Figure 3** (a) XRD $2\theta$-$\omega$ scans evolution of a superconducting sample immersed in water. The curves are vertically offset for clarity. (b) $\rho(T)$ curve versus immersion time. (c) A zoom-in view of (b). Insert of (c) shows superconducting transition temperature under different immersion time.

We then move from air to water. It is well-known that different cuprate superconductors display very large variation in their reactivity of water. For example, the famous $YBa_2Cu_3O_7$ is highly sensitive to water and water vapor [46]. In this respect, moisture in the air may be the culprit leading to the increase in resistivity. Therefore, it is meaningful to study the water stability of the superconductivity in nickelates. For this purpose, another superconducting film was soaked in de-ionized water at 20 °C. With increasing immersion time, the film diffraction peaks gradually weaken, broaden, and finally disappears (Figure 3(a)), revealing water at room temperature strongly degrades the structure of $Nd_{0.8}Sr_{0.2}NiO_2$. Here, five hours is a critical time. Before that, the diffraction peak intensity did not change significantly, while after that, the intensity began to decay obviously.

Correspondingly, this critical time is also reflected in the transport measurements. Figure 3(b) summarizes the resistivity evolution of film soaked in water. Superconductivity still exists within 2 hours of immersion. However, the sample lost its zero-resistance character when soaking for 5 hours, the resistivity is saturated at a finite value (0.5 mΩ·cm) at low temperatures, while the superconducting transition still exists. Once the strength of the diffraction peaks is weakened (corresponding to 8 hours), an insulating state emerges. Further prolonging the immersion time, i.e. more than 10 hours, would result in complete insulation ($\rho > 700$ Ω·cm at room temperature). Figure 3(c) displays a zoom-in view of Figure 3(b), focusing on the details of the superconducting transition regime. Within two hours, as the immersion time increases, the increase of resistivity is similar to the sample in air, except that the resistivity changes more drastically. Insert of Figure 3(c) illustrates the rapid change in $T_c$, suggesting obvious broadening effect in superconducting transition. Combined with the aging test in air and water, it is reasonable that the moisture is the main cause of the degradation of the sample in the air. In addition, it is also reasonable that degradation is a gradual process, which starts from the reaction of water/moisture and superconducting film at the surface. Oxygen and carbon dioxide in the air are secondary factors. Water seems to be the natural nemesis of high-$T_c$ superconductivity, and nickel-based superconductivity is no exception.

A simple surface degeneration model is proposed to explain the above phenomena based on two assumptions. The *first assumption* is that water/moisture could gradually decompose the sample from top (surface) to bottom. As illustrates in insert of Figure 2(c), due to the surface reaction with moisture, the actual effective thickness ($t_e$) of the superconducting film would gradually become thinner than original thickness ($t_0$). Since The film thickness was considered to be the default value ($t_0$) during testing, which means $t_e \leqslant t_0$. We assume that $t_e = t_0$ in the pristine measurement. In the subsequent exposure process, the surface of the sample gradually degrades, resulting in a gradual decrease in $t_e$. The *second assumption* is that the resistivity in $t_e$ part is equal to pristine resistivity. This assumption is reasonable since the degenerated layer on the surface is insulating, which would not contribute to the conduction. In addition, the trivial change in superconducting transition also support this hypothesis (insert of Figure 2(b)).

According to the resistivity relationship: $R = G \cdot \rho \cdot t_e^{-1}$, where $G$ is the geometric factor depends on the electrode position. The electrode position has not changed in the experiment; therefore, $G$ is a constant in all of these measurements. The resistance $R$ is then only proportional to $t_e^{-1}$, indicating the decrease in $t_e$ leads to increase in $R$. Since we used $t_0$ instead of $t_e$ by default in the measurement, resistivity is calculated by $\rho = G \cdot R \cdot t_0$. As a result, the resistivity $\rho$ is overestimated than actual $G \cdot R \cdot t_e$. In fact, the reduction in effective thickness $t_e$ leads to an increase in $R$. Since the change in thickness is ignored, dividing the resistance $R$ by the original thickness $t_0$ directly would result in an apparent increase in resistivity.

In turn, based on this model we can calculate the effective thickness $t_e$, which is the actual superconducting layer. In the pristine case, $R_{pristine} = G \cdot \rho \cdot t_0^{-1}$, while for degraded same film, $R_{degraded} = G \cdot \rho \cdot t_e^{-1}$. Therefore, $t_e / t_0 = R_{pristine} / R_{degraded} = \rho_{pristine}/\rho_{degraded}$. Our calculation shows after 47 days in ambient dry air the $t_e / t_0$ ratio is 0.95, indicating a 5% depth degeneration on the surface. It has been pointed out that the resistivity change versus exposure time has a relationship like $\rho(t) \propto t^{0.5}$. Therefore, based on the above model, if we take the resistivity change at 150 K as an example, we estimate that in order to reach 10% degeneration of $\rho_{pristine}$, it takes a total of 136 days, that is another 89 days after the experimentally determined 47 days. The predicted 10% degeneration depth is shown in the insert of Figure 2(a). In order to delay surface damage, a dry environment is preferred. Meanwhile, the effective thickness estimation for sample soaked in water are 93.5% (0.5 hour), 75% (2 hours) and 2.5% (5 hours). The first two value seems reasonable, while the effective thickness for 5 hours is too thin to be metallic. We have to point out that the degradation of the film in water may not be as uniform as in air due to the strong interaction with water. Actually, due to the strong reaction between water and thin film, the superconducting transition is obviously broadened (insert of Figure 3(c)) in several hours, indicating the strong interaction could influence the electrical properties of effective layers and make the degeneration depth deviate from surface degeneration model.

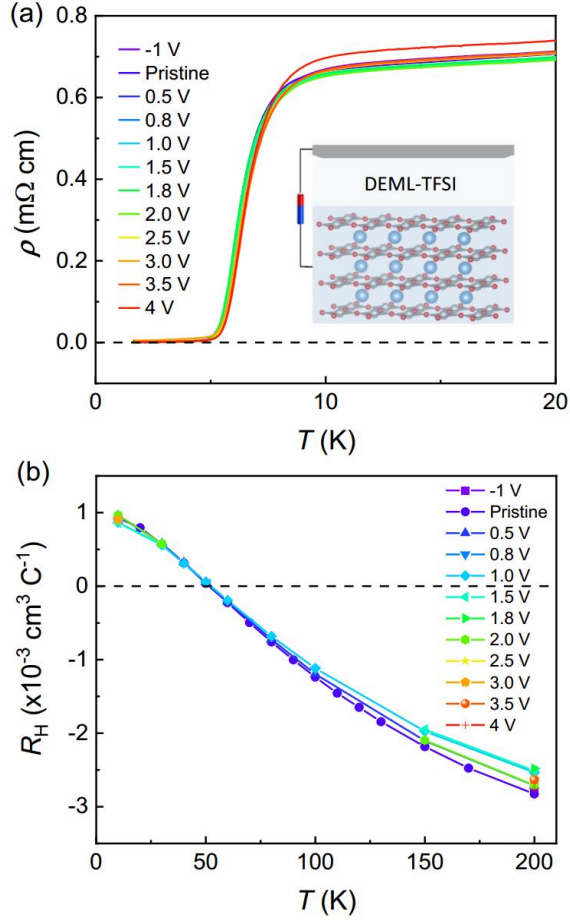

**Figure 4** (a) Temperature dependence of resistivity at multiple gate voltages in ionic-liquid gated $Nd_{0.8}Sr_{0.2}NiO_2$ film. The insert picture shows the schematic of the ionic-liquid-gated device. (b) Temperature-dependent Hall coefficient at multiple gate voltages in ionic-liquid gated $Nd_{0.8}Sr_{0.2}NiO_2$ film.

During the past few decades, ILG has shown great power in regulating carrier density (close to $8 \times 10^{14}$ cm$^{-2}$) in various materials [47-49]. In addition, the ILG also has potential to induce structural phase transition, i.e. the insertion of oxygen ions (negative voltage) or hydrogen ions (positive voltage) in the pristine material, such as in epitaxial $SrCoO_{2.5}$ thin films [35]. As a similar complex oxide whose apical oxygens are just removed by chemical reduction, infinite-layer structured $Nd_{0.8}Sr_{0.2}NiO_2$ might be prone to be re-oxidized back to perovskite $Nd_{0.8}Sr_{0.2}NiO_3$ or at least, $Nd_{0.8}Sr_{0.2}NiO_{2+\delta}$. Once this happened, the recovered apical oxygen atom would suppress the superconductivity in nickelates film and changes its temperature dependent resistivity curve.

With this motivation, we further explore the stability of infinite-layer $Nd_{0.8}Sr_{0.2}NiO_2$ superconducting thin film under even stronger perturbation conditions, such as direct electrochemical gating, to see if electrical properties can be effectively tuned. As shown in Figure 4 under the influence of ILG, the changes in both $T_c$ and $R_H$ are slight in a wide voltage window between -1 to 4 V. Since the upper limit of carrier modulation of ILG is ~ $10^{15}$ cm$^{-2}$, the observed insensitivity of Hall coefficient to applied electric field indicates that the intrinsic carrier density of nickelates is above this limit. In such a high carrier density case, an order of ~ $10^{15}$ cm$^{-2}$ control level provided by ILG is not sufficient to tune the carrier concentration in the film, leading to unchanged Hall coefficient.

In addition to the above explanation, the complex nature of infinite nickelates band structure might also contribute to the observed stability of $R_H$ under gating. In contrast to the well-established single orbital model developed for cuprates [50], superconducting nickelates demonstrate unique and more complex the band structure due to the simultaneous involvement of Nd $5d_{xy}$ electron pocked by self-doping and lifted Ni $3d_{x2-y2}$ orbital near the Fermi level [2, 51], thus the carrier density cannot be directly derived from Hall coefficient measurement. Although the three-dimensional rare earth $5d$ orbital hybridize with quasi-2D Ni $3d$ orbitals, it is still unclear what effect the $5d$ band may have on the strongly correlated Ni $3d$ band [52]. These facts increase the complexity of the data analysis. We infer that this multiband character leads to the failure of carrier extraction and need more exploration.

Additionally, Figure 4(a) and 4(b) also indicate the aforementioned structural phase transition does not appear in the current infinite-layer $Nd_{0.8}Sr_{0.2}NiO_2$ system. Based on previous reports, oxygenation and protonation could occur in both cuprates and iron pnictides [38, 53]. In those experiments, both 60nm $Pr_2CuO_{4\pm\delta}$ thin film and bulk iron pnictides (the 11 and the 122 structures) show phase transition in the voltage window between -1 to 4 V. In contrast, it is remarkably surprising that no obvious transport change can be observed in infinite nickelates in such a wide voltage interval. The unchanged structural and electrical properties of superconducting nickelate film under ILG suggests relative high energy barrier for hydrogen or oxygen ions evolution, indicating the superconducting nickelate film maintains robust against electrochemical environment.

# 4 Conclusion

In summary, we have prepared superconducting nickelates samples by pulsed laser deposition. XRD demonstrate the transition of $Nd_{0.8}Sr_{0.2}NiO_3$ to $Nd_{0.8}Sr_{0.2}NiO_2$ and temperature dependent resistivity reveals the onset $T_c$ about 8.0 ~ 11.1 K. By exposing the superconducting sample in air (20 °C, 35% RH) for 47 days, we find the superconductivity is quite stable with only a slight increase in resistivity. However, the superconductivity degrades quickly when sample is immersed in water. A simple surface degeneration model is developed to explain the increase in resistivity. These results indicate that superconductivity in nickelates is stable in dry air for long time, it is sensitive to moisture and water. Therefore, a low-humidity environment is preferable for their application. Further experiments demonstrate that ionic liquid gating has a little influence to the superconductivity and transport properties of the $Nd_{0.8}Sr_{0.2}NiO_2$ film, this in turn indicate that nickelate superconductor is very robust even in harsh electrochemical environment.

*This work is supported by NSFC of China (Grant No. 11774044, 52072059 and 11822411).*